\documentclass[12pt,a4paper]{article}      

\usepackage{amsthm,amsmath,amssymb,xcolor,sectsty,latexsym,mathtools,hyperref,mathrsfs}
\usepackage{slashed}  
\usepackage{bm}
\usepackage{epsfig}
\usepackage{dcolumn}
\definecolor{vermelho}{cmyk}{0,.88,.77,.40}
\usepackage{cite}
\usepackage{feynmp}
\numberwithin{equation}{section}
\usepackage[textwidth=6.3in,top=1.3in,bottom=1.3in]{geometry}
\usepackage{setspace}
\chapterfont{\color{vermelho}}
\usepackage{simplewick}

\newcommand{\be}{\begin{equation}}
\newcommand{\ee}{\end{equation}}
\newcommand{\beq}{\begin{equation}}
\newcommand{\eeq}{\end{equation}}
\newcommand{\ba}{\begin{eqnarray}}
\newcommand{\ea}{\end{eqnarray}}
\newcommand{\bef}{\begin{figure}}
	\newcommand{\eef}{\end{figure}}

\newcommand{\g}{\gamma}

\begin{document}
	
	\thispagestyle{empty}
	\begin{titlepage}
		\nopagebreak
		
		\title{  \begin{center}\bf  \Large Is Patience a Virtue? \\ Cosmic Censorship of Infrared Effects in de Sitter
		\vspace{1.5cm} 
\end{center} }
\vfill

\author{Ricardo Z. Ferreira$^{a}$\footnote{rferreira@icc.ub.edu}, ~ McCullen Sandora$^{b}$\footnote{mccullen.sandora@tufts.edu}~ and ~  Martin S. Sloth$^{c}$\footnote{sloth@cp3.sdu.dk}
}
		\maketitle
		
		\begin{center}
	\vspace{-0.7cm} 
	{\it  $^a$Departament de F\'isica Qu\`antica i Astrof\'isica i Institut de Ci\`encies del Cosmos}\\
	{\it  Universitat de Barcelona, Mart\'i i Franqu\`es, 1, 08028, Barcelona, Spain}\\
	\vspace{0.2cm}
	{\it  $^b$Institute of Cosmology, Department of Physics and Astronomy}\\
	{\it  Tufts University, Medford, MA 02155, USA}\\
	\vspace{0.2cm}
	{\it  $^c$CP$^3$-Origins, Center for Cosmology and Particle Physics Phenomenology}\\
	{\it  University of Southern Denmark, Campusvej 55, 5230 Odense M, Denmark}\\
	
\end{center}
	
		\begin{abstract}
		
While the accumulation of long wavelength modes during inflation wreaks havoc on the large scale structure of spacetime, the question of even observability of their presence by any local observer has lead to considerable confusion.  Though it is commonly agreed that infrared effects are not visible to a single sub-horizon observer at late times, we argue that the question is less trivial for a \emph{patient observer} who has lived long enough to have a record of the state before the soft mode was created. Though classically there is no obstruction to measuring this effect locally, we give several indications that quantum mechanical uncertainties censor the effect, rendering the observation of long modes ultimately forbidden.

	\end{abstract}
	
	\vfill
	\noindent \small Honorable Mention for the Gravity Research Foundation 2017 Awards for Essays on Gravitation

		\end{titlepage}
\thispagestyle{empty}
\section{Introduction}

De Sitter space is rather bedeviling to understand, which is unfortunate, given that it is, as far as we can tell, how our universe was born, and also how it will die.  Its very nature is rife with paradox: the gravitational effects of vacuum energy create a boundless amount of space, and yet, cruelly, this same process shields this unlimited bounty from each observer, confining us to the small backyard we call our observable universe.  Through its expansion it smoothes away any existing inhomogeneities, producing a large, flat spacetime, and yet as a result, minute disturbances are created, spoiling the perfect homogeneity.  This phenomenon muddles the notion of virtual and classical, as vacuum fluctuations are ripped far apart, and in the process decohere.  Each individual fluctuation has a very small amplitude, yielding a perturbative description most other branches of physics can only dream about.  Yet, if allowed to run long enough, these tiny affects may destroy our ability to describe the spacetime globally.

This is because long wavelength (soft) fluctuations accrue, warping the spacetime on very long length scales. The time it takes for this to occur, in analogy with the backreaction of Hawking radiation in the case of black hole evaporation, is the Page time of de Sitter, as has been argued \cite{Giddings:2011zd, Giddings:2011ze}.  This can be seen as follows: during every e-fold of de Sitter expansion, $\Delta t = 1/H$, two soft gravitons are
emitted at the horizon. After $N_e \sim M_p^2/H^2$ e-folds,
the number of quanta emitted rivals the horizon degrees of freedom giving rise to the de Sitter entropy $S\sim \text{Area}$.  In this regime, our standard descriptions must break down, and indeed they do: correlators diverge, such as $\langle\gamma_L^2\rangle\sim H^3t/M_p^2$, loop effects become the same order as tree level \cite{Giddings:2011zd, Giddings:2011ze}, and the quantum state of the system has vanishing overlap with the initial state \cite{1609.06318}.

But is this effect physical, or purely a deficiency of our description? This question has been addressed from many different sides (see \cite{Bartolo:2007ti, Giddings:2011zd, Giddings:2011ze} for an incomplete list of references).  Locally, this process is indistinguishable from a change in the background coordinates.  This has lead some to conjecture that infrared (IR) effects in de Sitter are completely unphysical gauge artifacts \cite{Urakawa:2010it}.

However, in certain circumstances this effect is measurable.  In de Sitter space information is not necessarily permanently lost when it recedes behind an observer's horizon.  If inflation ends, the modes that have left will eventually all be recovered.  Two satellites separated by vast differences will eventually be able to directly compare the changes in their coordinate charts induced by these long modes.  This has been called a \emph{meta-observer} \cite{Witten}, and is reliant on the nonlocal setup of the measurement.  The question becomes, can an analog of this process be carried out entirely within a single horizon?

In lieu of placing two measuring devices far apart in space, one may try to make two measurements far apart in time.  This defines what we call a {\it patient observer}, as it would take as long as a Page time to see any effect.  However, there is no {\it a priori} obstruction to creating a device capable of carrying out these measurements.

Or so it would seem.  Measurability is strongly challenged by quantum mechanical effects on the physical detectors themselves, as we find when considering different possible setups.  We cannot claim to have exhausted all possible measuring scenarios, but for the two examples we do show in detail, these effects are exactly below the fundamental threshold for what the devices can register.  This suggests a version of cosmic censorship forbidding an observer in de Sitter from measuring IR quantum gravitational effects.

\section{Particle production and gauge invariance}

As is well known, one can incorporate the effects a long mode locally through a coordinate transformation 
\begin{eqnarray}
k^2 \rightarrow e^{-\gamma_L}_{\,ij} k_i k_j \equiv \tilde{k}^2
\end{eqnarray}
where $\gamma_L$ is the soft tensor fluctuation.  Since this observer will still view themselves in flat space, this represents an \emph{asymptotic symmetry} of the de Sitter spacetime \cite{AnnNgStrom}.  As a consequence, one can associate a charge to this transformation \cite{[HintKhouryHui]} which encapsulates the change in the quantum state of the system \cite{1609.06318}. But this will have an effect on the short modes, which can be described as a Bogoliubov transformation \cite{ours}
\begin{eqnarray}
\phi_{\tilde{k}} = \alpha_k(\gamma_L) \, \phi_k + \beta_k(\gamma_L) \, \phi_k^* \, .
\end{eqnarray}
These coefficients can be readily computed, and the standard formula for the number of particles of momentum $k$ seen, $N_k =|\beta_k|^2$ applied. This yields \cite{ours}
\begin{eqnarray}
	N_k= \frac{\left< \g_L^2 \right>}{30} \, .
\end{eqnarray}
This means that, when the variance of the long mode becomes order unity, which happens at the Page time, a single extra particle (per mode) should be detected on top of the Gibbons-Hawking radiation \cite{Gibbons:1977mu}.

However, as we will discuss in the following section, for a physical detector, built out of actual matter, other quantum effects significantly alter this simple picture, and seem to forbid the possibility of such a measurement.

\section{Decoherence vs Patience}

We present two thought experiments on how one would go about measuring the effects of long modes, and each time we are stymied by the requirement that we build our machine out of physically realizable matter. This does not constitute a proof that a measurement device of sufficiently clever design cannot be envisioned, but it does give some indication that there may be a version of ``cosmic censorship'' at play, preventing one from making these measurements. This situation is reminiscent of \cite{DYSON:2013jra}, where it is argued that no machine capable of observing a single graviton may ever be constructed.

\begin{centering}
	\begin{figure*}[h]
		\centering
		\includegraphics[width=12cm]{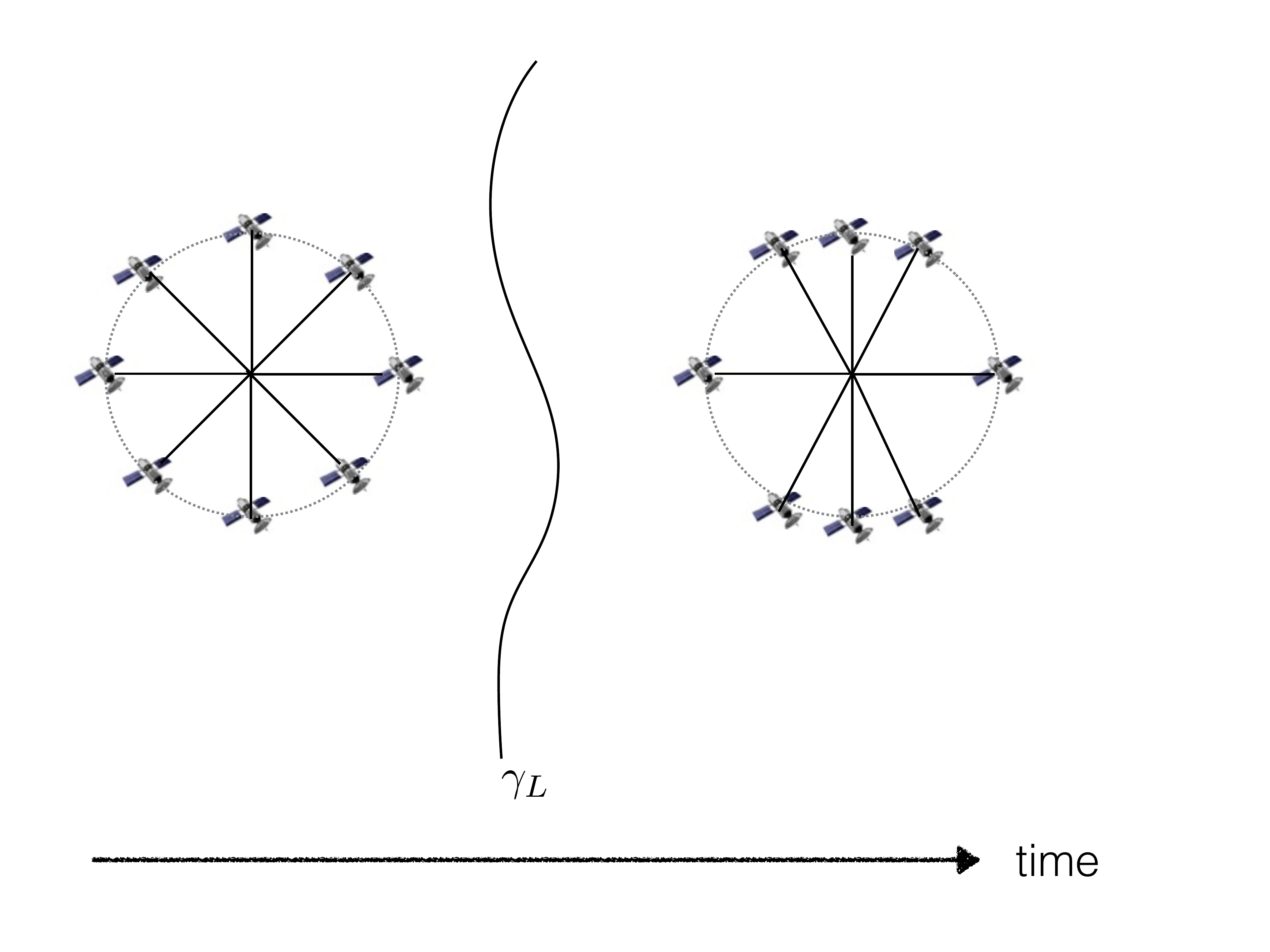}
		\caption{Example of a patient observer, an array of satellites radially connected like a carousel. When a soft mode is emitted, due to the shear stress it generates, the satellites move in the tangential direction.}
		\label{there}
	\end{figure*}
\end{centering}

The first example is an array of satellites connected radially to some concentric point through some wires, just like a carousel. The rigidity of the wires shields the radial expansion of the universe but allows for angular displacement of the satellites (see fig. \ref{there}). Alternatively we could think of the same array of satellites inside some rigid circular tube, like a hula hoop, where they would be allowed to move along the tube if some shear acts on them.

As a graviton passes through this setup, it induces a shear stress, and so displaces the satellites from their original angular positions. The relative change in the position is given by
\begin{eqnarray}
\left( \frac{\Delta x_\text{IR}}{x} \right)^2 \sim \left< \gamma_L^2 \right> \sim \frac{H^3 }{M_p^2} t.
\end{eqnarray}
Though this effect grows with time, to address its observability we need to compare it with the fundamental uncertainty in position coming from the Heisenberg uncertainty principle, $\Delta p \Delta x_q \gtrsim 1$. This implies 
\begin{eqnarray}
	\Delta x_q^2\gtrsim \frac{t}{m}\sim \frac{M_p^2}{m H^3}\left(\frac{\Delta x_{IR}}{x}\right)^2\sim\left(\frac{L_H}{r_s}\right)\left(\frac{L_H}{x}\right)^2\Delta x_\text{IR}^2
\end{eqnarray}
where $r_s$ is the Schwarzschild radius associated with the detector of mass $m$, and $L_H$ is the horizon size. Therefore, in order for the effect from long modes to be greater than the quantum uncertainty, either the detectors must be separated by a distance greater than the horizon, or they must be so massive that their Schwarzschild radii are larger than the horizon.  Thus, quantum effects prevent a local observer from measuring long modes in this way.

Another realization of a patient observer consists of comoving satellites exchanging electromagnetic signals, in hopes of measuring a relative time delay induced by the long modes through the redshift of light. Each satellite has its own clock and both are synchronized when the device is built. During the time the satellites are in causal contact they will measure the redshift and deduce from there either a change in time or in the Hubble constant through Hubble's law $z = H t$. The time shift induced by the soft gravitons can be read directly from the metric to be
\begin{eqnarray}
\Delta t_\text{IR} \sim \frac{\left< \gamma_L^2 \right>}{H} \sim  \frac{H^2 t}{M_p^2} \, .
\end{eqnarray}
However, in order for the clocks to have a precision $\Delta t$, they must have an energy uncertainty $\Delta E\gtrsim 1/\Delta t$.  Then, as pointed out in \cite{clocks}, physical clocks must interact with each other gravitationally, causing time dilation effects that will shift the time between clocks by an uncertain amount
\begin{eqnarray}
	\Delta t_\text{dilation}\sim\frac{\Delta E }{M_p^2 x} t \, . 
\end{eqnarray}
Comparing this to the effect of the long mode, we find
\begin{eqnarray}
\Delta t_\text{dilation}\sim \left(\frac{\Delta E}{H}\right)\left(\frac{L_H}{x}\right)\Delta t_{IR}
\end{eqnarray}
Therefore, when the physical requirements that the detector be able to fit inside the Hubble horizon and the minimal energy probed by the satellite cannot be less than the Hubble rate are imposed, we find that these again exactly forbid the effect to be measured.

We have performed various extensions of these simple systems, as well as several other novel strategies for measurement, and each time we have run into the same problem. We leave for future work if the case of inflation, where $H$ changes in time, brings any improvement.

\section{Conclusion}

What we have shown here is that although infrared effects in de Sitter become physical if an observer is able to perform some nonlocal measurement, when considering specific realizations we always face the same kind of censorship: {\it quantum effects on the detectors are of exactly the right size to forbid the measurement of near-soft gravitons}.  

This leads us to conjecture that this is an indication of a fundamental limitation to the observability of infrared effects in de Sitter space. This points to a tantalizing sort of quantum-geometric consistency, that even though the geometry of spacetime `fuzzes out' on such scales, it does so in such a way that it is just below the threshold of detectability for any local observer.  Patience may be a virtue, but it is not one that the universe holds in high regard.\\

{\bf \noindent Acknowledgements:}
RZF is supported by ERC Starting Grant HoloLHC- 306605. MSS is supported by Villum Fonden grant 13384. CP3-Origins is partially funded by the Danish National Research Foundation, grant number DNRF90.

\bibliographystyle{JHEP}
\bibliography{gravityessay.bib}

\end{document}